\documentclass[runningheads]{llncs}
\usepackage[T1]{fontenc}
\usepackage{graphicx,verbatim}
\usepackage[ruled,vlined]{algorithm2e}  
\usepackage{amsmath, amsfonts} 
\usepackage{hyperref}
\usepackage{booktabs}
\begin{document}
\title{Skeleton-Guided Diffusion Model for Accurate Foot X-ray Synthesis in Hallux Valgus Diagnosis}
\titlerunning{Skeleton-Guided Diffusion Model for Accurate Foot X-ray Synthesis}
%

\author{
Midi~Wan\inst{1} \and
Pengfei~Li\inst{2} \and
Yizhuo~Liang\inst{1} \and
Di~Wu\inst{3} \and
Yushan~Pan\inst{4} \and
Guangzhen~Zhu\inst{5} \and
Hao~Wang\inst{1}
}

\authorrunning{Wan et al.}

\institute{
Xidian University, Xi'an, China\\
\email{wanghao@xidian.edu.cn}
\and
Beijing Tongren Hospital, Capital Medical University, Beijing, China
\and
Norwegian University of Science and Technology, Trondheim, Norway
\and
Xi'an Jiaotong–Liverpool University, Suzhou, China
\and
Chongqing Jiaotong University, Chongqing, China
}

\maketitle              
\begin{abstract}

Medical image synthesis plays a crucial role in providing anatomically accurate images for diagnosis and treatment. Hallux valgus, which affects approximately 19\% of the global population, requires frequent weight-bearing X-rays for assessment, placing additional strain on both patients and healthcare providers. Existing X-ray models often struggle to balance image fidelity, skeletal consistency, and physical constraints, particularly in diffusion-based methods that lack skeletal guidance. We propose the Skeletal-Constrained Conditional Diffusion Model (SCCDM) and introduce KCC, a foot evaluation method utilizing skeletal landmarks. SCCDM incorporates multi-scale feature extraction and attention mechanisms, improving the Structural Similarity Index (SSIM) by 5.72\% (0.794) and Peak Signal-to-Noise Ratio (PSNR) by 18.34\% (21.40 dB). When combined with KCC, the model achieves an average score of 0.85, demonstrating strong clinical applicability. The code is available at \url{https://github.com/midisec/SCCDM}.


\end{abstract}
\section{Introduction}

Foot X-ray imaging is a critical tool for diagnosing deformities such as Hallux Valgus (HV). However, its high cost and limited accessibility hinder its use for widespread screening\cite{cai2023global}. Patients with HV often require repeated weight-bearing foot X-rays, both before and after surgery, to ensure accurate assessment and monitoring\cite{bernasconi2022bosch,kanatli2020effect}. This necessity further exacerbates the burden on patients and healthcare systems.

While deep learning models have shown promise in synthesizing X-ray images from natural light, they frequently struggle to accurately capture fine bone structures, such as fractures and lesions. This limitation significantly restricts their clinical applicability. In the field of cross-modal medical image synthesis, recent advancements have focused on improving the realism and logical consistency of generated images. Key strategies include the use of rigid datasets\cite{huang2021coarse}, multi-scale modules\cite{zhou2020hi,dalmaz2022resvit}, attention mechanisms\cite{li2024hybrid,QIMS125898,zhang2024bcswinreg}, and physical priors\cite{jin2023physics,nemirovsky2024explicit} to ensure that synthesized images adhere to physical constraints.

Despite these advancements, no existing research specifically addresses the synthesis of foot X-ray images. Moreover, there is currently no standardized evaluation method to assess the anatomical accuracy of such synthesized images. Lately, Denoising Diffusion Probabilistic Model (DDPM) have gained significant attention for their strong generative capabilities. Due to their iterative nature, they have been successfully applied to medical image synthesis\cite{pinaya2022brain,yoon2023sadm}, segmentation\cite{wu2024medsegdiff,wu2024medsegdiffV2}, reconstruction\cite{pan2023diffuseir,han2024physics}, and classification\cite{yang2023diffmic,yang2025diffmic}.

However, to the best of our knowledge, there is currently no diffusion model designed for foot X-ray image synthesis. We train the DDPM in a diff-domain, which differs from traditional image-space training. The diff-domain is defined as the pixel-wise difference between the target X-ray image and the conditional natural light image. Specifically, we encode foot images captured under natural light as conditional inputs and compute the diff-domain during the diffusion process to guide the model in predicting noise at each stage. Unlike existing models, our method incorporates a self-attention branch and a multi-scale structural module into the diffusion sampling process. By integrating both local skeletal keypoints and global foot structure information, our method enhances structural consistency in the generated images. As a result, our approach is capable of generating high-fidelity X-ray images with fine details while maintaining realistic skeletal and joint structures. We also introduce a novel evaluation metric, KCC, which is based on a pretrained foot keypoint detection model. We compare our method against standard synthesis methods. 

In this paper, we propose the Skeleton-Constrained Conditional Diffusion Model (SCCDM) in a diff-domain setting for foot X-ray synthesis. To address the constrained issue of ensuring logical skeletal structures in hallux valgus diagnosis, we introduce multi-scale feature extraction and self-attention mechanisms. We further propose a novel evaluation metric, KCC, to assess skeletal integrity in the generated X-ray images.

\section{Methods}
\begin{figure}
\centering
\includegraphics[width=0.9\textwidth]{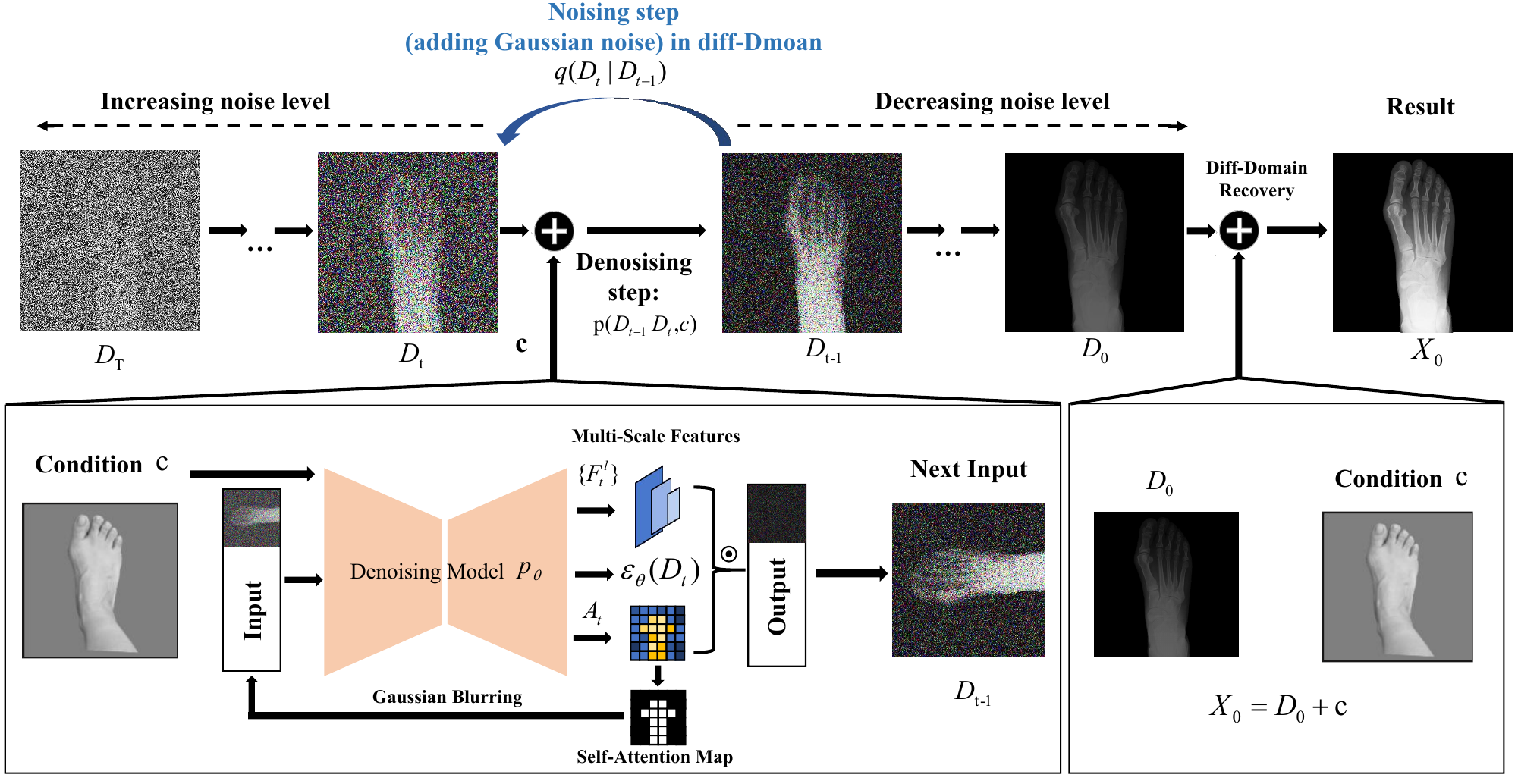}
\caption{Overview of the Skeletal-Constrained Conditional Diffusion Model. The diffusion process operates in the diff-domain, defined as the pixel-wise difference between the target X-ray image and the conditional natural light image. } \label{overview}
\end{figure}

Figure~\ref{overview} illustrates the overview of our method. The diffusion process in the diff-domain is detailed in Sect.\ref{sec:overview}. The denoising network integrates multi-scale modules and attention mechanisms while the operation of KCC is introduced in Sect.\ref{sec:kcc}.

\subsection{Denoising Diffusion Generative Models} \label{sec:overview}
In the field of medical image synthesis, a typical task is the transformation between different imaging modalities. This process requires ensuring that the generated images are not only of high visual quality but also anatomically reasonable. Our research is based on diffusion models, proposing a detail-preserving image-to-image translation approach that effectively learns foot skeletal features and generates anatomically coherent X-ray images. Our method follows the formulation of DDPM as given in \cite{ho2020denoising,nichol2021improved}. In Algorithm~\ref{alg:refined_sag_sampling}, we present the overall workflow of our approach.

\begin{figure}
\centering
\includegraphics[width=0.9\textwidth]{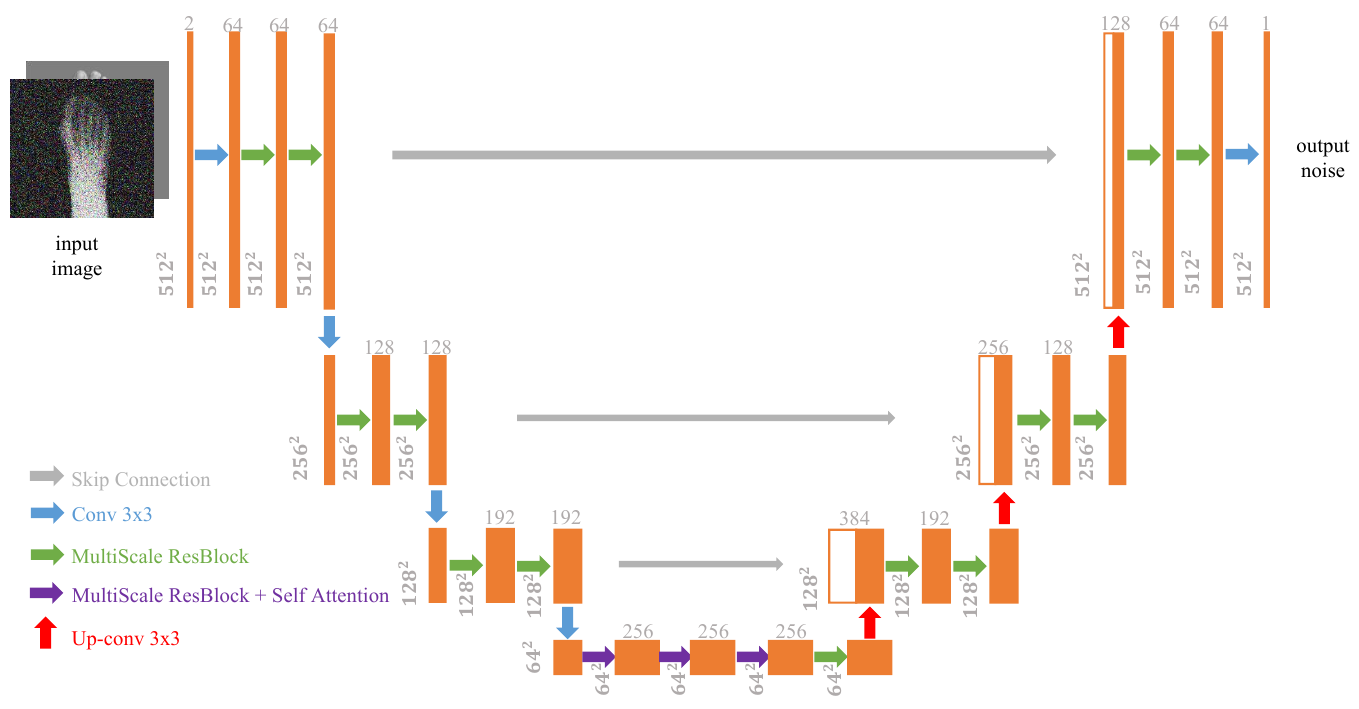}
\caption{U-Net Architecture for Diffusion Models with Multi-Scale and Self-Attention Modules for Noise Prediction.} \label{unet}
\end{figure}

  The general idea of diffusion models is to iteratively transform an image $\mathbf{x}$ into a series of noisy images $\{\mathbf{x}_0, \mathbf{x}_1, ..., \mathbf{x}_T\}$ by adding small amounts of Gaussian noise at each step. The noise level of an image $\mathbf{x}_t$ increases gradually as $t$ progresses from 0 to $T$. The U-net based $\epsilon_\theta$ model (Figure~\ref{unet}) is trained to predict $\epsilon_\theta$ from $\mathbf{x}_t$ and t according to a predefined noise schedule. During training, the ground truth $\epsilon$ is known, and the model is optimized using a mean squared error loss. During inference, we start from a random Gaussian noise $\mathbf{x}_T \sim \mathcal{N}(0, I)$ and iteratively predict $\mathbf{x}_{t-1}$ until reaching $\mathbf{x}_0$. This denoising process allows us to generate high-quality synthetic images. The forward diffusion process $q$ is defined as:
\begin{equation}
q(\mathbf{x}_t | \mathbf{x}_{t-1}) := \mathcal{N} (\mathbf{x}_t; \sqrt{1 - \beta_t} \mathbf{x}_{t-1}, \beta_t I).
\end{equation}
By recursively applying this noise formulation, we obtain:
\begin{equation}
\mathbf{x}_t = \sqrt{\bar{\alpha}_t} \mathbf{x}_0 + \sqrt{1 - \bar{\alpha}_t} \boldsymbol{\epsilon}, \quad \boldsymbol{\epsilon} \sim \mathcal{N}(0, I).
\end{equation}

\noindent
where $\alpha_t := 1 - \beta_t$ and $\bar{\alpha}_t := \prod_{s=1}^{t} \alpha_s$. The denoising process $p_\theta$ is trained to learn the reverse transition:

\begin{equation}
p_\theta(\mathbf{x}_{t-1} | \mathbf{x}_t) := \mathcal{N} (\mathbf{x}_{t-1}; \mu_\theta(\mathbf{x}_t, t), \Sigma_\theta(\mathbf{x}_t, t)).
\end{equation}

The output of the U-Net is denoted as $\epsilon_\theta$, and the training loss is:

\begin{equation}
\mathcal{L} := \|\boldsymbol{\epsilon} - \epsilon_\theta(\sqrt{\bar{\alpha}_t} \mathbf{x}_0 + \sqrt{1 - \bar{\alpha}_t} \boldsymbol{\epsilon}, t)\|^2_2, \quad \boldsymbol{\epsilon} \sim \mathcal{N}(0, I).
\end{equation}

To obtain $\mathbf{x}_{t-1}$, the standard DDPM formulation follows:

\begin{equation}
\mathbf{x}_{t-1} = \sqrt{\bar{\alpha}_{t-1}} \left( \frac{\mathbf{x}_t - \sqrt{1 - \bar{\alpha}_t} \epsilon_\theta(\mathbf{x}_t, t)}{\sqrt{\bar{\alpha}_t}} \right) + \sqrt{1 - \bar{\alpha}_{t-1} - \sigma_t^2} \epsilon_\theta(\mathbf{x}_t, t) + \sigma_t \boldsymbol{\epsilon}.
\end{equation}

\noindent
where \( \sigma_t = \sqrt{(1 - \bar{\alpha}_{t-1}) / (1 - \bar{\alpha}_t)} \sqrt{1 - \bar{\alpha}_t / \bar{\alpha}_{t-1}}. \)


\subsubsection{Skeletal-Constrained.} \label{sec:skeletal_constrained}
Our framework refines the standard diffusion process by incorporating multi-scale feature fusion and self-attention guidance. In the image synthesis phase, Given a condition image $\mathbf{c}$, we initialize the latent variable $\mathbf{d}_T \sim \mathcal{N}(0, I)$ and iteratively refine it through $T$ steps. At each step $t$, the diffusion model $p_\theta$ predicts the noise estimate $\epsilon_\theta$, the variance $\Sigma_t$, multi-scale feature maps $\{F_t^l\}$, and the attention map $A_t$. The multi-scale feature fusion module aggregates $\{F_t^l\}$ to construct a refinement mask $\mathbf{M}_t$, while the attention-based gating mechanism generates $\mathbf{G}_t$ based on a threshold $\psi$, $\psi = 0.5$. To enhance spatial consistency, the estimated clean image $\tilde{\mathbf{d}}_0$ is computed using:

\begin{equation}
\tilde{\mathbf{d}}_0 = (\mathbf{d}_t - \sqrt{1 - \bar{\alpha}_t} \epsilon_\theta) / \sqrt{\bar{\alpha}_t}.
\end{equation}

The multi-scale refinement is applied as:

\begin{equation}
\mathbf{d}_t^{\text{ref}} = (1 - \mathbf{M}_t) \odot \mathbf{d}_t + \mathbf{M}_t \odot \tilde{\mathbf{d}}_t.
\end{equation}

The self-attention-guided refinement ensures that regions of high activation are selectively updated:

\begin{equation}
\mathbf{d}_t^{\text{attn}} = (1 - \mathbf{G}_t) \odot \mathbf{d}_t^{\text{ref}} + \mathbf{G}_t \odot \tilde{\mathbf{d}}_t.
\end{equation}

The generated image is obtained by:

\begin{equation}
\mathbf{x}_0 = \mathbf{d}_0 + \mathbf{c}.
\end{equation}

\begin{algorithm}[H]
\caption{Refined Sampling with Multi-Scale and Self-Attention}
\label{alg:refined_sag_sampling}
\KwIn{Condition image $\mathbf{c}$, diffusion model $p_\theta$, diffusion steps $T$, noise schedule functions $\alpha_t$, $\beta_t$, $\bar{\alpha}_t$.}
\KwOut{Generated image $\mathbf{x}_0$}
\SetAlgoNlRelativeSize{0.4}
Initialize $\mathbf{d}_T \sim \mathcal{N}(0, \mathbf{I})$, $\psi = 0.5$\;
\For{$t = T$ down to $1$}{
    \If{$t > 1$}{Sample $\mathbf{z} \sim \mathcal{N}(0, \mathbf{I})$}
    \Else{$\mathbf{z} = \mathbf{0}$}
    
    $\epsilon_\theta, \Sigma_t, \{F_t^l\}, A_t \gets p_\theta(\mathbf{d}_t, \mathbf{c}, t)$\;
    $\mathbf{M}_t \gets \text{FuseMultiScale}(\{F_t^l\})$\;
    $\mathbf{G}_t \gets 1(A_t > \psi)$\;
    $\tilde{\mathbf{d}}_0 \gets (\mathbf{d}_t - \sqrt{1-\bar{\alpha}_t}\epsilon_\theta)/\sqrt{\bar{\alpha}_t}$\;
    $\tilde{\mathbf{d}}_0 \gets \text{GaussianBlur}(\tilde{\mathbf{d}}_0)$\;
    $\tilde{\mathbf{d}}_t \gets \sqrt{\bar{\alpha}_t}\tilde{\mathbf{d}}_0 + \sqrt{1-\bar{\alpha}_t}\epsilon_\theta$\;
    
    $\mathbf{d}_t^{\text{ref}} \gets (1 - \mathbf{M}_t) \odot \mathbf{d}_t + \mathbf{M}_t \odot \tilde{\mathbf{d}}_t$\;
    $\mathbf{d}_t^{\text{attn}} \gets (1 - \mathbf{G}_t) \odot \mathbf{d}_t^{\text{ref}} + \mathbf{G}_t \odot \tilde{\mathbf{d}}_t$\;
    
    $\boldsymbol{\mu}_t \gets \frac{1}{\sqrt{\alpha_t}} (\mathbf{d}_t^{\text{attn}} - \frac{\beta_t}{\sqrt{1 - \bar{\alpha}_t}} \epsilon_\theta )$\;
    $\mathbf{d}_{t-1} \sim \mathcal{N}(\boldsymbol{\mu}_t, \Sigma_t)$\;
}
$\mathbf{x}_0 \gets \mathbf{d}_0 + \mathbf{c}$\;
\Return $\mathbf{x}_0$\;
\end{algorithm}

\subsection{Comprehensive Evaluation Metrics} \label{sec:kcc}
When evaluating the generated images at the pixel level, Structural Similarity Index Measure (SSIM), Peak Signal-to-Noise Ratio (PSNR), and Mean Absolute Error (MAE) are three commonly used metrics that measure the similarity between generated and real images from different numerical perspectives. 
These three metrics complement each other: SSIM emphasizes structural similarity, PSNR focuses on noise levels, and MAE directly measures error. For bone structure assessment, two widely used metrics - Fréchet Inception Distance (FID) and Learned Perceptual Image Patch Similarity (LPIPS) - evaluate image quality from different perspectives. 

\begin{equation}
FID = ||\mu_r - \mu_g||^2 + \text{Tr}(\Sigma_r + \Sigma_g - 2(\Sigma_r \Sigma_g)^{(1/2)})
\end{equation}

\noindent
where $\mu_r, \Sigma_r$ and $\mu_g, \Sigma_g$ represent the mean and covariance of the feature distributions for real and generated images, respectively. 

\begin{equation}
LPIPS(x, y) = \sum_{l} \frac{1}{H_l W_l} \sum_{h,w} || \phi_l (x)_{hw} - \phi_l (y)_{hw} ||^2
\end{equation}

\noindent
where $l$ represents the layer index of the pre-trained deep learning feature extractor, and $H_l, W_l$ denote the height and width of layer $l$. Unlike traditional pixel-based metrics, LPIPS focuses on perceptual similarity and structural coherence. 

To comprehensively evaluate the anatomical plausibility of generated X-ray images at foot keypoint locations, we propose the Keypoint Confidence-Completeness (KCC) metric. This metric quantifies the confidence and completeness of keypoint localization, allowing for a detailed analysis of the recognition reliability and completeness of skeletal keypoints. Using a pre-trained Diff3dhpe skeletal keypoint detection model, denoted as $D(\cdot)$, the model outputs $M$ keypoints and their corresponding confidence scores $\{(k_i(\hat{x}), c_i(\hat{x})) | i = 1, ..., M\}$. Based on this, we define the comprehensive evaluation function as follows:

\begin{equation}
\mathcal{E}(\hat{x}, D(\cdot), \tau) = \left\{
\begin{aligned}
D(\hat{x}) &= \{(k_i(\hat{x}), c_i(\hat{x})) | i = 1, ..., M\} \\
S(\hat{x}) &= \frac{1}{M} \sum_{i=1}^{M} c_i(\hat{x}) \\
KCC (\hat{x}; \tau) &= S(\hat{x}) \times \frac{1}{M} \sum_{i=1}^{M} 1 (c_i(\hat{x}) \geq \tau)
\end{aligned}
\right.
\end{equation}

\noindent
where $S(\hat{x})$ is the average confidence score of all detected keypoints. $KCC (\hat{x}; \tau)$ represents the keypoint confidence-completeness score, which combines the average confidence with the proportion of keypoints detected above a confidence threshold $\tau$, $\tau$=$0.65$. $k_i(\hat{x})$ represents the predicted coordinates of the $i$-th keypoint in the generated image, and $c_i(\hat{x})$ is the confidence score of the $i$-th keypoint. $M$ is the total number of detected skeletal keypoints, max($M$) = $10$.

\section{Experiments}
The hyperparameters of the SCCDM model are as follows: $T = 1000$ sampling steps, Adam optimizer with a learning rate of $1e-5$, and a batch size of 2. The first layer consists of 64 channels, with one attention head applied at a resolution of 64. The channel multiplier is $[1, 2, 3, 4]$, and each layer contains 2 residual blocks with a dropout rate of 0.3. The diffusion process is defined by $\beta_1 = 1e-4$ and $\beta_T = 0.02$, and the model has a total of 37,539,809 parameters. Training is performed on two Nvidia RTX 4090 GPUs over a period of seven days. The implementation is based on PyTorch 2.4.1, and the Nat2XFoot dataset is used for evaluating our method.

\subsubsection{Nat2XFoot.} 
This dataset consists of paired foot images in natural light and X-ray modalities, collected from 150 patients at a Beijing hospital between June and August 2024. Each patient provided one foot image, standardized to 512 × 512 pixels. Foot region segmentation was performed using the Segment Anything Model\cite{kirillov2023segment}, and image registration techniques aligned the images. After expert verification, 145 high-quality pairs were retained, split into 115 training, 15 validation, and 15 test images. All images were intensity-normalized with pixel values scaled to [0, 1].

 \begin{figure}
 \centering
\includegraphics[width=0.9\textwidth]{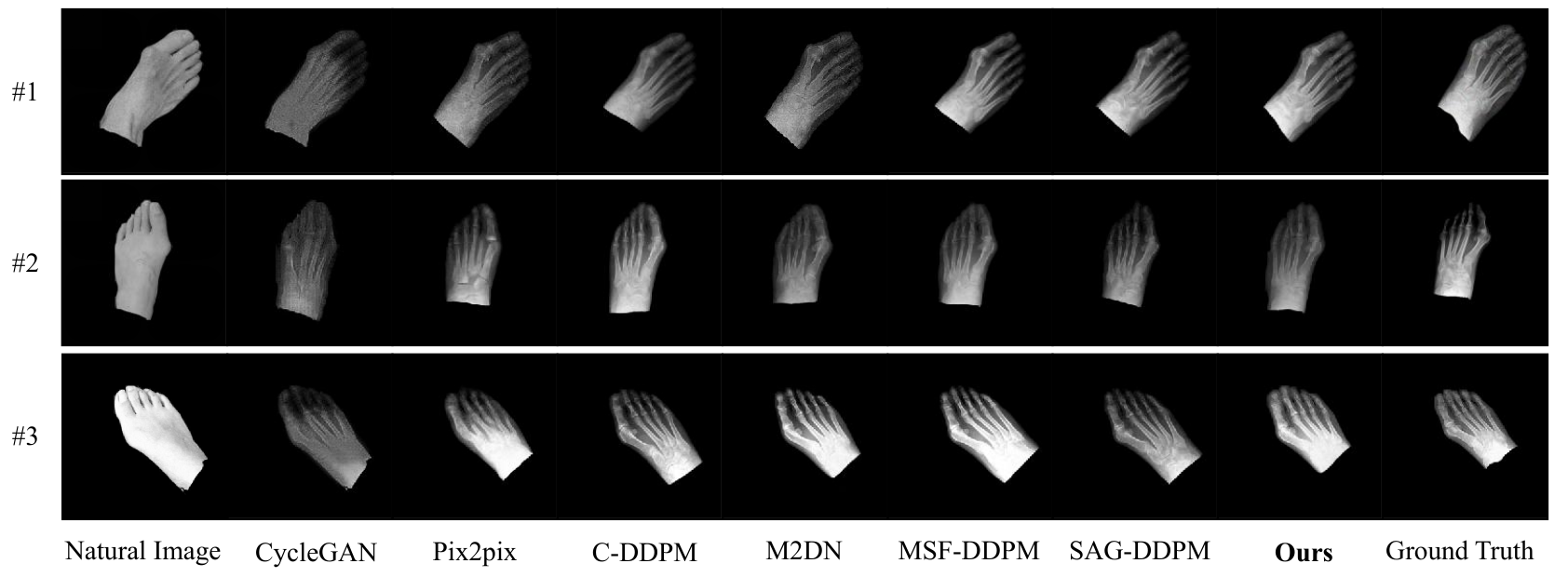}
\caption{Comparison of X-ray Image Synthesis Methods for Hallux Valgus: Natural Images Translated to X-ray Images.} \label{result}
\end{figure}

\begin{table}[h]
    \small  
    \centering
    \caption{Quantitative Performance Comparison of X-ray Image Synthesis Methods.}
    \label{tab:performance_comparison}
    \renewcommand{\arraystretch}{1.1} 
    \setlength{\tabcolsep}{4pt} 
    \begin{tabular}{lcccccc}
        \toprule
        {Method} & {SSIM} & {PSNR} & {MAE} & {FID} & {LPIPS} & {KCC} \\
        \midrule
        CycleGAN \cite{zhu2017unpaired}  & 0.713 $\pm$ 0.03 & 17.13 $\pm$ 0.8 & 0.0667 & 274.15 & 0.2147 & 0.44 \\
        pix2pix \cite{isola2017image}   & 0.745 $\pm$ 0.02 & 17.92 $\pm$ 0.6 & 0.0524 & 167.22 & 0.1897 & 0.62 \\
        C-DDPM \cite{peng2023generating}     & 0.756 $\pm$ 0.02 & 18.18 $\pm$ 0.7 & 0.0496 & 146.74 & 0.1814 & 0.65 \\
        DDPM-MAR \cite{karageorgos2024denoising}   & 0.753 $\pm$ 0.02 & 17.75 $\pm$ 0.6 & 0.0509 & 156.21 & 0.1724 & 0.68 \\
        M2DN \cite{meng2024multi}      & 0.762 $\pm$ 0.02 & 18.31 $\pm$ 0.7 & 0.0529 & 143.49 & 0.1754 & 0.72 \\
        MSF-DDPM \cite{yang2024fontdiffuser}  & 0.785 $\pm$ 0.02 & 20.21 $\pm$ 0.8 & 0.0398 & 135.56 & 0.1767 & 0.78 \\
        SAG-DDPM \cite{hong2023improving}  & 0.754 $\pm$ 0.02 & 17.87 $\pm$ 0.6 & 0.0528 & 141.63 & 0.1867 & 0.74 \\
        \textbf{SCCDM (ours)}  & \textbf{0.794 $\pm$ 0.02} & \textbf{21.40 $\pm$ 0.9} & \textbf{0.0326} & \textbf{132.34} & \textbf{0.1673} & \textbf{0.85} \\
        \bottomrule
    \end{tabular}
\end{table}

\begin{table}[h]
    \small
    \centering
    \caption{Ablation study on SCCDM. We evaluate the effect of each module by removing one at a time.}
    \label{tab:ablation_study}
    \renewcommand{\arraystretch}{1.1} 
    \setlength{\tabcolsep}{4pt} 
    \begin{tabular}{lcccccc}
        \toprule
        {Ablation Setting} & {SSIM} & {PSNR} & {MAE} & {FID} & {LPIPS} & {KCC} \\
        \midrule
        SCCDM w/o Diff-domain  & 0.775 & 18.91 & 0.0473 & 147.21 & 0.1660  & 0.75 \\
        SCCDM w/o Multi-scale  & 0.743 & 17.42 & 0.0503 & 150.44 & 0.1742 & 0.69 \\
        SCCDM w/o Self-attention  & 0.768 & 19.03 & 0.0453 & 144.02 & 0.1739 & 0.78 \\
        SCCDM  & 0.794  & 21.40 & 0.0326 & 132.34 & 0.1673 & 0.85 \\
        \bottomrule
    \end{tabular}
\end{table}

\section{Results and Discussion}

To evaluate our method, we compare SCCDM with several cross-modal image synthesis methods, including CycleGAN\cite{zhu2017unpaired}, Pix2pix\cite{isola2017image}, C-DDPM\cite{peng2023generating}, DDPM-MAR\cite{karageorgos2024denoising}, M2DN\cite{meng2024multi}, MSF-DDPM\cite{yang2024fontdiffuser}, and SAG-DDPM\cite{hong2023improving}. In Fig. \ref{result}, we show three typical natural images of hallux valgus patients from the Nat2XFoot dataset alongside the corresponding X-ray images generated by each method. Our approach consistently generates clinically acceptable X-ray images that preserve the foot’s joints and skeletal structure in line with anatomical details. In contrast, CycleGAN and Pix2Pix occasionally distort the image’s structure, leading to incomplete or misaligned bone and joint positions, which undermines the reliability of the results. HV requires calculating bone angles, and incorrect skeletal information (KCC < 0.75) impacts clinical performance. DDPM-based methods, however, produce more complete reconstructions of bones and joints, but the M2DN model sometimes introduces excessive skeletal details, resulting in unnatural X-ray images. We hypothesize that this is due to M2DN overfitting, especially in fine detail enhancement, which detracts from the biological accuracy of the foot anatomy. In comparison, SCCDM excels at maintaining consistent structural integrity for bones and joints while preserving contextual details. Only SAG-DDPM produces accurate joint and skeletal details, successfully reconstructing fine features while maintaining structural coherence. Table \ref{tab:performance_comparison} provides a quantitative comparison of the methods, evaluating image quality, structural accuracy, detail retention, and similarity to real X-ray images. SAG-DDPM achieves a KCC of 0.74, suggesting that the attention-based mechanism may capture the relationships between joints and bones. However, it still struggles to fully replicate the complex anatomical details in a single scale, leading to some discrepancies in bone structure and joint alignment. On average, our method improves by 5.72\% in SSIM and 18.34\% in PSNR compared to other methods.

Ablation studies in Table \ref{tab:ablation_study} show that removing any module degrades both image quality and anatomical accuracy, with the multi-scale module being the most critical. Without the diff-domain module leading to a loss of fine structural details essential for distinguishing bone boundaries. Excluding the multi-scale module causes the most severe deterioration in skeletal alignment, reducing SSIM to 0.743 and KCC from 0.85 to 0.69. This decline suggests that the generated X-ray images fail to maintain clinically relevant skeletal relationship. Misalignment in bone structures directly impacts the accuracy of HV diagnosis. Similarly, removing self-attention lowers SSIM to 0.768 and increases FID from 132.34 to 144.02, indicating disrupted global coherence and joint articulation.

\section{Conclusion}
We propose a novel skeleton-guided image synthesis method that combines multi-scale feature extraction with self-attention to address the challenges of generating accurate X-ray images for HV patients. Introducing the concept of diff-domain in the diffusion process enhances structural fidelity and detail retention. We also introduce the KCC metric to assess keypoint localization confidence and completeness. Experimental results show that SCCDM outperforms existing methods in image quality, structural consistency, and detail reconstruction, achieving an SSIM of 0.794 and PSNR of 21.40, outperforming state-of-the-art techniques. Our code is
available at \url{https://github.com/midisec/SCCDM}.

%
%
%
%

\bibliographystyle{splncs04}
\bibliography{references}







\end{document}